# Benchmarking Geant4 photonuclear process model for the photo-induced reaction of deformed nuclei in the GDR region

P. D. Khue[1], P. V. Cuong[2], D.L. Balabanski[3] , L. X. Chung[1], D.V. Thanh[4], D. T. K. Linh[1], and

L. T. Anh[1*]

[1]*Institute for Nuclear Science and Technology, VINATOM, 179 Hoang Quoc Viet, Hanoi, Vietnam.*
[2]*Hanoi Irradiation Center, VINATOM, 2PXV+473, QL32, Minh Khai, Hanoi, Vietnam.*
[3]*Extreme Light Infrastructure - Nuclear Physics/Horia Hulubei National Institute for R&D in Physics and Nuclear Engineering, Bucharest-Magurele, Romania*
[4]*Institute of Theoretical and Applied Research, Duy Tan University, Hanoi, Vietnam.*
.

**Abstract:**

The Geant4 photonuclear process is benchmarked by comparing it with experimental data to verify the ability of the Geant4 toolkit to simulate the photo-induced reaction on deformed nuclei in the Giant Dipole Resonance (GDR) region. The simulation results are compared with experimental data of the deformed nuclei ($^{153}$Eu, $^{160}$Gd, $^{165}$Ho, and $^{186}$W) targets in terms of both the average neutron energies and the photonuclear cross-sections. The agreement between the calculated results of the Geant4 photonuclear process model and the experimental measurements is analyzed.

## I. INTRODUCTION

Photonuclear reactions are essential for a wide range of fundamental research and applications. They play a crucial role in the synthesis of nuclei in the Universe and offer researchers a very useful tool to probe the atomic nucleus because of their selectivity and the reaction mechanism's model-independence [1]. In the last decades, the rapid development of advanced photon sources with high intensity and quality in conjunction with cutting-edge detector technology has led to the revival of the study of photonuclear reactions in the laboratory. Facilities, such as ELI-NP in Romania [2], HIγS in the USA [3], NewSUBARU in Japan [4], and SLEGS in China [5], that provide high-intensity, quasi-monoenergetic, tunable, and polarized photon beams facilitate many kinds of photonuclear studies: nuclear photoabsorption, nuclear resonance fluorescence, and photonuclear reactions.  Especially, photonuclear reactions of deformed nuclei in the GDR region always draw interest from the scientific community. Nuclear deformation reflects the collective motion that comes from the interaction between valence nucleons and shell structure. Studying the nuclear reactions of deformed nuclei in the GDR region offers not only a sensitive test of the hydrodynamic theory [6, 7] but also a way to extract clues to the understanding of nuclear structure [8].

The Monte Carlo method has proven useful for the interpretation of experimental data, testing the theory as well as designing experimental setups. Some widely used Monte Carlo codes are Fluka [9, 10], Phits [11], and Geant4 [12, 13, 14]. Of these codes, Geant4 has become one of the most ubiquitous codes due to its flexibility.

The Geant4 photonuclear process model does two things: generates the final state of outgoing particles and calculates the total photonuclear inelastic cross-section. For the incoming photons with energy below 200 MeV, G4LowEGammaNuclearModel is used to generate the final state based on preequilibrium and equilibrium reaction mechanisms. The Geant4 toolkit offers two built-in photonuclear cross-section models that are handled by G4PhotoNuclearCrossSection (G4PNXS) and G4GammaNuclearXS (G4GNXS) classes, respectively.

In the previous study, we benchmarked the Geant4 photonuclear process model with encouraging results. Nonetheless, the benchmark primarily focused on the comparison of experimental photoneutron cross-section with simulation results and was conducted mainly for spherical nuclei. Furthermore, in terms of the cross-section model, only G4PNXS was validated [15].

The work reported here is a nice addition to previous work with the main focus on the photonuclear reaction of the deformed nuclei in the GDR region. Especially, not only the experimental photoneutron cross-sections but also the measured average neutron energy spectra are used to benchmark the Geant4 photonuclear process model. G4GNXS is also under consideration in this work. Moreover, we also introduce here a new interface that allows using the output of Talys - a computer code system for the analysis and prediction of nuclear reactions [16]- in Geant4 simulation for the photonuclear process. This study is also useful guidance for applying Geant4 in supporting photonuclear studies.

## II. EXPERIMENTAL REVIEW

The results of the simulations are benchmarked by comparisons with experimental data measured by B. L. Berman et al. [8] at Livermore Linear Accelerator Facility, and by T. Shizuma et al. [17] at the National Institute of Advanced Industrial Science and Technology in Japan.

The measurements by B. L. Berman et al. were made with the quasi-monoenergetic photon beam obtained from the annihilation in-flight of fast positrons, and with an efficient paraffin- and- BF3-tube $4\pi$ neutron detector for neutron-multiplicity counting. The ring-ratio technique was deployed to determine the average energy of the photo-ejected neutrons for both single- and double-photoneutron events. Photonuclear cross-sections for $^{153}$Eu, $^{160}$Gd, $^{165}$Ho, and $^{186}$W were experimentally determined as a function of photon energy from 8 to 29 MeV. These cross-sections included $\sigma[(\gamma,n)+(\gamma,pn)]$, $\sigma[(\gamma,2n)+(\gamma,p2n)]$, and $\sigma(\gamma,3n)$. The photon energy resolution ranged from less than 300 keV at the lowest energies to 400 keV at the highest energies, and the data were acquired at intervals of 300 keV or less. The uncertainty of absolute cross sections was determined to be at most 7% [8].

In the second experiment under consideration, T. Shizuma et al. used quasi-monochromatic photon beams from laser Compton scattering (LCS) with average energies ranging from 7.3 to 10.9 MeV to estimate the experimental $^{186}$W($\gamma$,n) cross-sections. The neutrons were detected using sixteen 3He proportional counters. The LCS photon beam had an energy resolution of 10% in FWHM. In the meantime, 5.9–8.3% was estimated to be the entire systematic uncertainty for measured cross-sections [17].

## III. SIMULATION INTRODUCTION

### A. Photonuclear reaction mechanism in the GDR region

In the GDR region, the photonuclear reaction is a two-step process that involves the formation of a compound nucleus [1]. In the first stage, the target nucleus absorbs the incident photon and enters an excited state. After reaching equilibrium, the compound nucleus deexcites through the evaporation of gammas, neutrons, protons, or other light ions. The deexcitation process occurs independently of the formation of the compound nucleus.

The deexcitation of a compound nucleus is simulated according to the Weisskopf-Ewing [18] or the Hauser-Feshbach [19] methods. The main difference between the two methods is that the Hauser-Feshbach theory considers the conservation of the total angular momentum and parity, whereas the Weisskopf–Ewing approach, which is used by Geant4, neglects it.

### B. Simulation model

#### 1. Geant4 low-energy gamma-nuclear final stage model

In Geant4, the sampling of the final stage for a reaction induced by photons at energies below 200 MeV is handled by the low-energy gamma-nuclear final stage model (implemented under the G4LowEGammaNuclearModel class) which comprises preequilibrium and equilibrium mechanisms [12, 20, 21]. Up until the nuclear system reaches equilibrium, the precompound stage of the nuclear reaction is taken into consideration.

Based on the Griffin semiclassical exciton model [22], the Geant4 preequilibrium considers the pre-compound nucleus to be composed of an undisturbed nucleus and a system of excitons conveying the excitation energy and momentum. During this stage, there is a competition between emissions of particles (protons, neutrons, deuterons, tritium, and helium nuclei) and transitions to states with a different number of excitons, $\Delta n$ (where $\Delta n$ = +2, -2, 0). When the number of excitons is approximately equal to the equilibrium number of excitons, the transition from the preequilibrium phase to the compound state is performed.

After the preequilibrium phase, the residual nucleus supposedly enters an equilibrium stage and becomes a compound nucleus with excitation energy $E^*$. The subsequent decay of the compound nucleus occurs through the emissions of particles, fragments, or photons if the excitation energy is greater than the separation energy. Within the Geant4 framework, the compound nucleus can deexcite [21] through:

- the evaporation of nucleons and light particles handled by a hybrid model consisting of two semi-classical methods: (1) Weisskopf-Ewing [18] standard evaporation model for neutrons and light-charged ions with $Z \leq 2$; (2) Generalize Evaporation Model [23] for heavier particles up to $^{28}$Mg,
- the evaporation of photons based on the tubulated E1, M1, and E2 transition probabilities for discrete emission, and based on E1 giant dipole resonance strength distribution for continuous emission,
- Bohr-Wheeler semi-classical model [24] for the fission process,
- the emission of multi-fragments calculated by the statistical multi-fragmentation model from [25],
- or Fermi break-up based on [25].

2. *Geant4 built-in cross-section for photonuclear reactions*

The G4PNXS class parameterizes photonuclear cross-sections that cover all incident photon energies up to the hadron generation threshold. There are five energy regions in the parameterization [20], each of which corresponds to the dominant physical process, as listed below:

- The GDR region can range from 10 MeV up to 30 MeV, depending on the nucleus.
- The second region known as the quasi-deuteron, which has small cross-sections with a broad and low peak, spans from around 30 MeV to the pion threshold.

- The next three regions correspond to the Δ region (from pion threshold to 450 MeV), the Roper resonance region (from roughly 450 MeV to 1.2 GeV), and the Reggeon-Pomeron region (upward from 1.2 GeV), respectively.

Since the Geant4.11.0 version, a new photonuclear cross-section model (G4GNXS) based on the IAEA Evaluated Photonuclear Data Library [26] at GDR energy region from 0 MeV to 130 MeV has been introduced [27]. The IAEA Evaluated Photonuclear Data Library contains evaluated photonuclear data for 219 isotopes stored in ENDF-6 format files. Either G4PhysicsLinearVector or G4PhysicsFreeVector vector containers were used to retrieve and interpolate the data from database files and put them in the simulation process.

*3. IS2020 a Geant4-based code for photonuclear studies*

In the previous work [15], we introduced the implementation of a Geant4-based code IS2020 to simulate photonuclear reactions where G4LowEGammaNuclearModel is utilized for sampling the final stage and G4PNXS is for calculation of photonuclear cross-section.

In this work, IS2020 is updated to include G4GNXS as an alternative option for photonuclear cross-section. Moreover, a new interface is also added to IS2020 to construct a Talys-based cross-section model for the Geant4 simulation process, based on the output file of Talys code [16]. The interface is designed to be able to read directly the Talys output file with the format "reaction.tot" where Talys stores its calculated reaction cross-section.

### C. Simulation condition setup

In this study, we used the IS2020 code running on the 11.2.0 version of Geant4 to simulate the photonuclear reactions on $^{153}$Eu, $^{160}$Gd, $^{165}$Ho, and $^{186}$W targets. The number of monoenergetic incident photons is set to $10^{10}$ for the simulation. The size, shape, and composition of the targets for the simulation are consistent with those in the experiment. Three simulation configurations were considered:

- Using G4PNXS.
- Using G4GNXS.
- Using Talys-based cross-section model (Talys-basedXS) with reaction cross-section files calculated by Talys1.9 with default parameters.

All configurations use G4LowEGammaNuclearModel for sampling the final stage. The energies of the monoenergetic photon beam varied from 8 MeV to 30 MeV with a step of 0.5

MeV. In each simulation, the characteristics of reactions, such as outgoing particles and their kinematics, were stored in a "root" file to be post-analyzed by ROOT packages [28].

## IV. RESULTS

To verify the ability of the photonuclear process model in Geant4 to regenerate the products of photo-induced reactions on deformed nuclei targets, the simulation results are compared with the experimental data taken from [8, 17].

### A. Average photoneutron Energies

Figure 1 demonstrates the simulated energy spectra of single- and double-photoneutron events for $^{153}$Eu. The spectra shift to the right when the energy value of the incident photon beam increases. In our simulations with $10^{10}$ primary photons, we seldom observed the (γ,pn) and (γ,p2n) reactions. This implies that the corresponding energy spectra are mainly contributed by the (γ,n) and (γ,2n) reactions. The similar results were obtained for $^{160}$Gd, $^{165}$Ho, and $^{186}$W. The average neutron energies were determined as the weighted mean of the corresponding energy spectra.

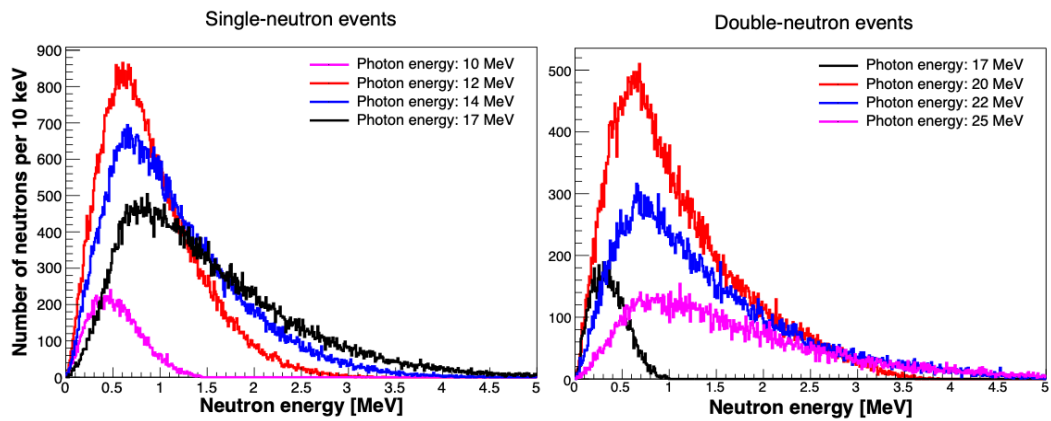

*Figure 1: Demonstration for simulated neutron energy spectra of single-neutron events (left) and double-neutron events (right) for $^{153}$Eu with different photon energies.*

Figure 2 presents the dependence of the average neutron energies on the photon energy for $^{153}$Eu, $^{160}$Gd, $^{165}$Ho, and $^{186}$W targets respectively. The simulation results are displayed in blue color, whilst the black circles represent the experimental data. The simulated average energies of emitted neutrons are handled by the Geant4 low-energy gamma-nuclear final stage model and are independent of the used photonuclear cross-section model. Thus, only results obtained with G4GNXS are shown in comparison with experimental data. The average energy spectra of single- and double-photoneutron events are drawn in separate subplots.

As confirmed in [8], the experimental data had relatively poor statistics which can be observed by the scatter of the data points. However, it is sufficient to be used for benchmarking the Geant4 low-energy gamma-nuclear final stage model. In all cases, the average neutron energy rose rapidly against the energy values of the photon beams far above the reaction thresholds for both experimental data and simulation data. Additionally, simulations also show that the average energies of single-photoneutron events are mainly contributed by the (γ,n) reactions rather than (γ,pn) reactions (see text for Figure 1). It holds for single-photoneutron events as well with the main contribution coming from (γ,2n) reactions. Generally speaking, within the poor statistics of the measured data points, the simulation results are in decent agreement with the experimental data.

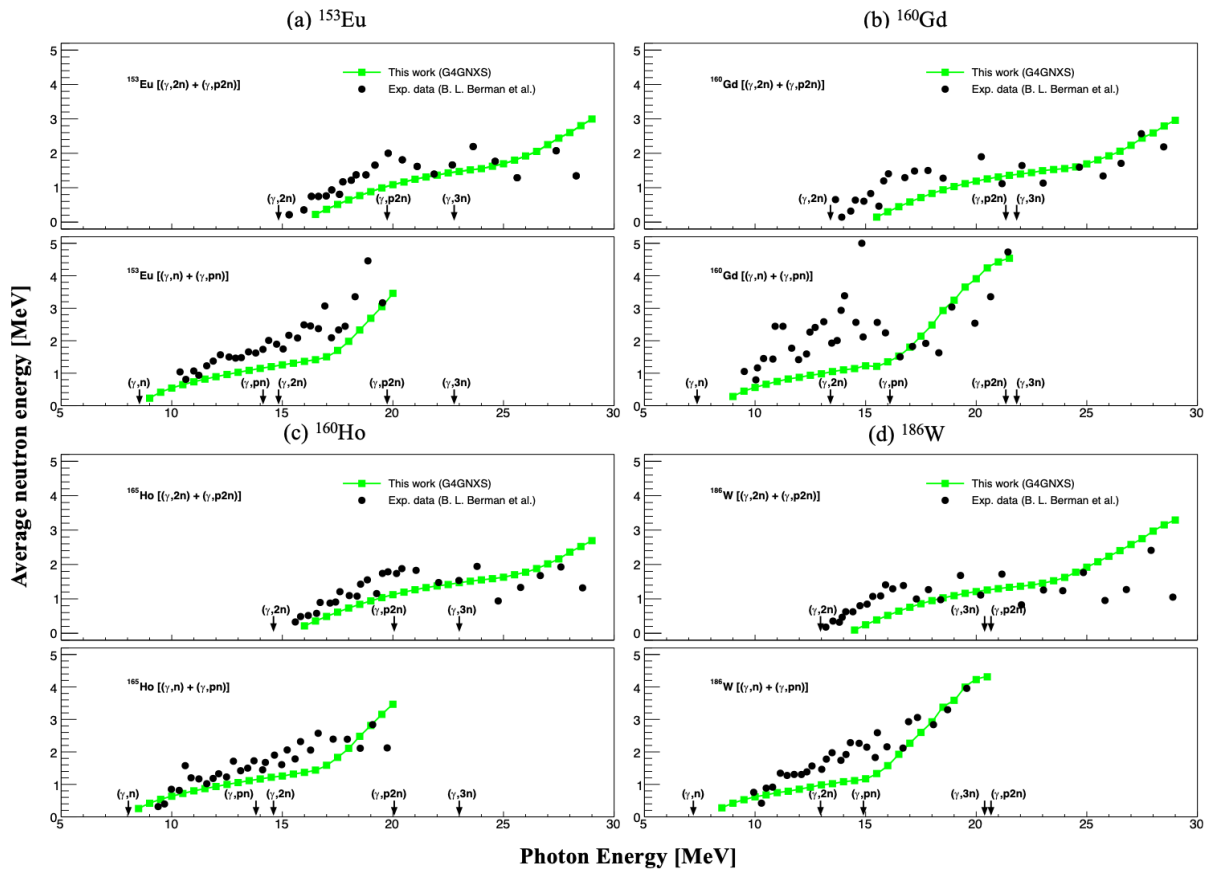

Figure 2: The dependence of average neutron energies of the [(γ,2n) + (γ,p2n)] and [(γ,n) +(γ,pn)] events on photon energy for: $^{153}$Eu (a), $^{160}$Gd (b), $^{165}$Ho (c), and $^{186}$W (d). The arrows indicate the photoneutron threshold values taken from [8].

## B. Photoneutron Cross-sections

$^{153}$Eu, $^{160}$Gd, $^{165}$Ho, and $^{186}$W are well-known deformed nuclei. Experimental measurements indicated that their photonuclear cross-sections contained two separate peaks in the GDR region.

Figure 3 presents the comparison of total photoneutron cross-sections for $^{153}$Eu, $^{160}$Gd, $^{165}$Ho, and $^{186}$W deformed nuclei between simulations and experimental data, where the black circles represent the experimental data points. As mentioned in Section. III.B, three simulation configurations corresponding to three cross-section models are considered. The total photoneutron cross-sections, $\sigma_t = \sigma[(\gamma,n) +(\gamma,pn)+ (\gamma,2n) + (\gamma,p2n) + (\gamma,3n))]$, obtained with these configurations are plotted in red, green, and blue, respectively.

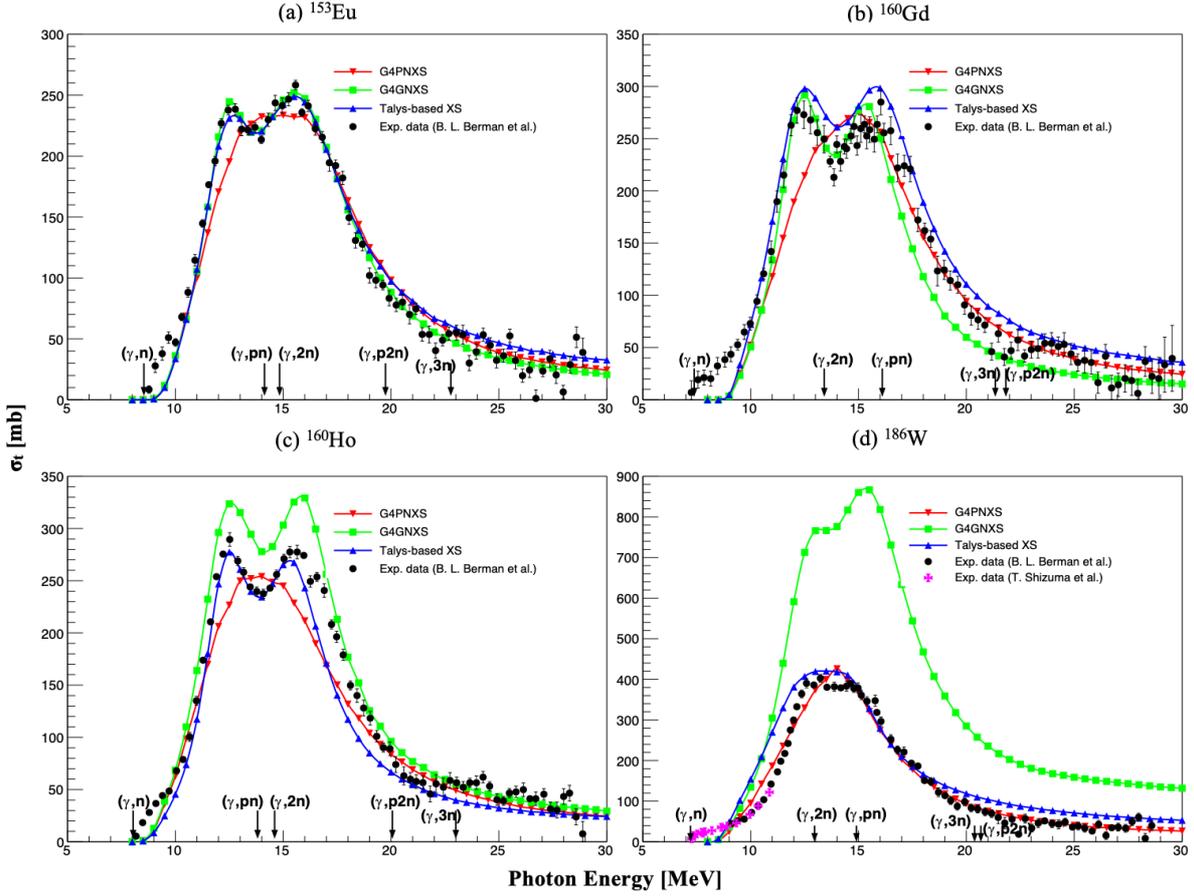

*Figure 3: The dependence of total photoneutron cross-section $\sigma_t$ on photon energy for: $^{153}$Eu (a), $^{160}$Gd (b), $^{165}$Ho (c), and $^{186}$W (d). The arrows indicate the photoneutron threshold values taken from [8].*

For the case of $^{153}$Eu, a good agreement between the measured data and results obtained with G4GNXS and Talys-basedXS. Meanwhile, G4PNXS only describes the experimental data well in the energy range from approximately 17 MeV onwards. Especially, G4PNXS cannot reproduce the two-peaked giant resonance as observed for measured data.

For the case of results for $^{160}$Gd, G4PNXS describes the measured data in the energy range above approximately 16 MeV better than G4GNXS and Talys-basedXS. Nevertheless, like the case of $^{153}$Eu, G4PNXS failed to regenerate two separate maxima of the giant resonance.

Generally speaking, G4GNXS and Talys-basedXS describe the experimental data relatively well.

Figure 3c displays the simulation results and the experimental data for $^{165}$Ho. Again, we observe that G4PNXS fails to reproduce the double-humped giant resonance characteristic of prolate-deformed nuclei. Both G4GNXS and Talys-basedXS describe well the splitting of the giant resonance. However, the result generated by G4GNXS is higher than the measured data in the double-humped region. In general, Talys-basedXS regenerates the measured data better than G4GNXS.

The simulated total photoneutron cross-sections for $^{186}$W are shown in Figure 3d. Along with the experimental data taken from [8], the experimental data measured by T. Shizuma et al. [17] (plotted in pink color) are also included. There is a big difference between the simulation results obtained with G4GNXS and the measured data. G4GNXS generates cross-sections approximately many times larger than the experimental values, though it can describe the splitting of the giant resonance. In the meantime, the results of the G4PNXS and Talys-basedXS describe the measured data better than G4GNXS.

Looking at the region of a few MeVs above the (γ,n) threshold in Figure 3, it can be said that the Geant4 photonuclear process model fails to describe the cross-sections above the particle evaporation threshold for all considered deformed nuclei.

## V. DISCUSSION AND CONCLUSION

In this work, the Geant4 simulations were performed for the photonuclear reaction of deformed nuclei in the GDR region to obtain the average neutron energies and the photoneutron cross-sections. The simulation results are compared with the existing experimental data [8, 17] for $^{153}$Eu, $^{160}$Gd, $^{165}$Ho, and $^{186}$W nuclei. The G4LowEGammaNuclearModel model is used in the simulation for generating the outgoing channels and their kinematics. For photonuclear cross-section, the performance of G4PNXS, G4GNXS, and a Talys-based model is analyzed.

In the GDR region, the main mechanism is the deexcitation of the compound nucleus through the evaporation of particles (mainly gammas and neutrons). Within the Geant4 framework, the kinematics of evaporated particles are sampled by the Weisskopf-Ewing method [18]. For the kinematics of photo-ejected neutrons, the G4LowEGammaNuclearModel model describes reasonably the trend and the value of the average energy. Bearing in mind that experimental data had relatively poor statistics due to the limitation of experimental conditions [8]. Therefore, the outcome of the comparison is encouraging.

For the case of the average energies for $^{160}$Gd (see in Figure 2b), B. L. Berman et al. [8] stated that there was seemingly a sharp dip in the measured average neutron energy spectrum starting at the (γ, pn) threshold (16.1 MeV). They assumed that the dip, given the high Coulomb barrier for protons, may be explained by the increase of the (γ, pn) reaction, which will add low-energy neutrons to the single-photoneutron events. However, our simulations show that due to a small cross-section, the contribution of (γ, pn) reactions to the calculation of average neutron energies can be ignored. Therefore, we assume that the so-called dip was merely the scattering of the data points due to the statistical limitations of the $^{160}$Gd ring-ratio data.

For the photoneutron cross-section, the G4PNXS model is incapable of reproducing the double-humped giant resonance characteristic of prolate-deformed nuclei. Meanwhile, G4GNXS regenerates the double-humped shape in all studied cases. G4GNXS describes relatively well the measured photoneutron cross-sections of $^{153}$Eu, $^{160}$Gd. However, its results for $^{165}$Ho and $^{186}$W are higher than the experimental data. Especially the discrepancy is remarkably high for $^{186}$W. After investigation, we found that the problem originates from the database behind G4GNXS. As mentioned in Section B, the G4NXS database was built from the IAEA Evaluated Photonuclear Data Library [26]. The exact reason, why extracted data has this problem requires further investigation. We reported this issue to Geant4 collaboration, and some efforts have been initiated to study the issue for finding a solid solution.

The Talys-based cross-section model, in general, describes decently well the experimental data of all studied nuclei.

In summary, this work shows that the G4LowEGammaNuclearModel model can reproduce well the experimental average neutron energies in the GDR region for deformed nuclei. We used the simulation results to re-interpret the measured data of the average energies for $^{160}$Gd. None of the two Geant4 native cross-section models can accurately reproduce the measured data of all four studied deformed nuclei. Except for the $^{165}$Ho and $^{186}$W cases, G4GNXS describes the experimental data better than G4PNXS. One can consider using the Talys-based cross-section model as an alternative. The Geant4 photonuclear process model still has room for improvement in predicting more accurately the photo-induced reactions above the particle evaporation threshold.

**ACKNOWLEDGMENT**

This work was supported by the National Foundation for Science and Technology of Vietnam (NAFOSTED) under Grant No.103.04-2021.76. Regarding G4GNXS, L.T. Anh would like to thank Dr. V. Ivantchenko (CERN, CH1211, Geneva 23, Switzerland) for his useful discussion.## REFERENCES

[1] A. Zilges, D. L. Balabanski, J. Isaak and N. Pietralla, Photonuclear reactions—From basic research to applications, Prog. Part. Nucl. Phys. 122, 103903 (2022).
[2] S. Gales, K. Tanaka, D. L. Balabanski and et al, The extreme light infrastructure—nuclear physics (ELI-NP) facility: new horizons in physics with 10 PW ultra-intense lasers and 20 MeV brilliant gamma beams, Rep. Prog. Phys. 81, 094301 (2018).
[3] A. Tonchev, M. Boswell, C. Howell et al., The high intensity γ-ray source (HIγS) and recent results, Nucl. Instrum. Methods Phys. Res. B 241, 170 (2005).
[4] S. Amano and et al., Several-MeV-ray generation at NewSUBARU by laser Compton backscattering, Nucl. Instrum. Methods Phys. Res. A 602, 337 (2009).
[5] Z. Pan and et al., Design and dynamic studies for a compact storage ring to generate gamma-ray light source based on Compton backscattering technique, Phys. Rev. Accel. Beams. 22, 040702 (2019).
[6] M. Goldhaber and E. Teller, On Nuclear Dipole Vibrations, Phys. Rev. 74, 1046 (1948).
[7] H. Steinwedel, J. Hans D. Jensen and P. Jensen, Nuclear Dipole Vibrations, Phys. Rev. 79, 1019 (1950).
[8] B. L. Berman, M. A. Kelly, R. L. Bramblett and et al., Giant Resonance in Deformed Nuclei: Photoneutron Cross Sections for $^{153}$Eu, $^{160}$Gd, $^{165}$Ho and $^{186}$W, Phys. Rev. C 185, 1576 (1969).
[9] G. Battistoni, The FLUKA code: Description and benchmarking, AIP Conf. Proc. 896, 31 (2007).
[10] G. Battistoni and et al., Overview of the FLUKA code, Ann. Nucl. Energy 82, 10 (2015).
[11] T. Sato and et al., Features of Particle and Heavy Ion Transport code System (PHITS) version 3.02, J. Nucl. Sci. Technol. 55, 684 (2018).
[12] J. Allison, K. Amako, J. Apostolakis and et al., Recent developments in Geant4, Nucl. Instrum. Methods Phys. Res. A 835, 186 (2016).
[13] J. Allison, K. Amako, J. Apostolakis and et al., Geant4 Developments and Applications, IEEE Trans. Nucl. Sci. 53, 270 (2006).
[14] S. Agostinelli, J. Allison, K. Amako and et al., Geant4 - A Simulation Toolkit, Nucl. Instrum. Meth. A 506, 250 (2003).
[15] L. T. Anh, P. V. Cuong, H. T. Thao and et al., Implementation of a Geant4-based code using low-energy gamma-nuclear final state model for photonuclear studies, Nucl. Instrum. Methods Phys. Res. A, 1027, 166285 (2022).
[16] A. Koning and D. Rochman, Modern Nuclear Data Evaluation with the TALYS Code System, Nucl. Data Sheets 113, 2841 (2012).
[17] T. Shizuma, H. Utsunomiya, P. Mohr and et al., Photodisintegration cross section measurements on $^{186}$W, $^{187}$Re, and $^{188}$Os: Implications for the Re-Os cosmochronology, Phys. Rev. C 72, 025808 (2005).
[18] V. E. Weisskopf and D. H. Ewing, On the Yield of Nuclear Reactions with Heavy Elements, Phys. Rev. 57, 472 (1940).
[19] W. Hauser and H. Feshbach, The Inelastic Scattering of Neutrons, Phys. Rev. 87, 366 (1952).
[20] Geant4 Collaboration, Geant4 Manual: Physics Reference.
*Available at: https://geant4.web.cern.ch/ support/user_documentation.*
[21] J. M. Quesada, V. A. Ivanchenko and et al., Recent Developments in Pre-Equilibrium and De-Excitation Models in Geant4, Prog. Nucl. Sci. Technol. 2, 936 (2011).
[22] J. J. Griffin, Statistical Model of Intermediate Structure, Phys. Rev. Lett. 17, 478 (1966).
[23] S. Furihata, K. Niita, S. Meigo, Y. Ikeda and F. Maekawa, The GEM Code - A Simulation Program for the Evaporation and Fission Process of an Excited Nucleus, JAERI-Data/Code 2001-015, Japan Atomic Energy Research Institute (JAERI), 2001.
[24] N. Bohr and J. A. Wheeler, The Mechanism of Nuclear Fission, Phys. Rev. 56, 426 (1939).
[25] J. P. Bondorf, A. S. Botvina, A. S. Iljinov, I. N. Mishustin and K. Sneppen, Statistical multifragmentation of nuclei, Phys. Rep. 257, 133 (1995).